# Evaluation of silicon carbide as a divertor armor material in DIII-D H-mode discharges

T. Abrams[1], S. Bringuier[1], D.M. Thomas[1], G. Sinclair[1], S. Gonderman[1], L. Holland[1], D.L. Rudakov[2], R.S. Wilcox[3], E.A. Unterberg[3], and F. Scotti[4]

[1] General Atomics, San Diego, CA 92121, United States of America
[2] University of California San Diego, La Jolla, CA 92093, United States of America
[3] Oak Ridge National Laboratory, Oak Ridge, TN 37831, United States of America
[4] Lawrence Livermore National Laboratory, Livermore, CA 94550, United States of America

E-mail: abramst@fusion.gat.com

**Abstract**

Silicon carbide (SiC) represents a promising but largely untested plasma-facing material (PFM) for next-step fusion devices. In this work, an analytic mixed-material erosion model is developed by calculating the physical (via SDTrimSP) and chemical (via empirical scalings) sputtering yield from SiC, Si, and C. The Si content in the near-surface SiC layer is predicted to increase during D plasma bombardment due to more efficient physical and chemical sputtering of C relative to Si. Silicon erosion from SiC thereby occurs primarily from sputtering of the enriched Si layer, rather than directly from the SiC itself. SiC coatings on ATJ graphite, manufactured via chemical vapor deposition, were exposed to repeated H-mode plasma discharges in the DIII-D tokamak to test this model. The qualitative trends from analytic modeling are reproduced by the experimental measurements, obtained via spectroscopic inference using the S/XB method. Quantitatively the model slightly under-predicts measured erosion rates, which is attributed to uncertainties in the ion impact angle distribution, as well as the effect of edge-localized modes. After exposure, minimal changes to the macroscopic or microscopic surface morphology of the SiC coatings were observed. Compositional analysis reveals Si enrichment of about 10%, in line with expectations from the erosion model. Extrapolating to a DEMO-type device, an order-of-magnitude decrease in impurity sourcing, and up to a factor of 2 decrease in impurity radiation, is expected with SiC walls, relative to graphite, if low C plasma impurity content can be achieved. These favorable erosion properties motivate further investigations of SiC as a low-Z, non-metallic PFM.

## 1. Introduction

Low-Z plasma-facing components (PFCs) have long been recognized as instrumental in achieving the highest possible plasma performance in magnetic fusion energy (MFE) devices [1]. For example, the low-Z walls of DIII-D are composed of graphite and carbon fiber composite (CFC) tiles and have enabled many recent successes in its steady-state, advanced tokamak (AT) and transient control programs [2]. Low-Z walls can be achieved in fusion devices either by using an intrinsically low-Z material or by active replenishment on top of a high-Z surface, thereby preventing erosive wear or other damage to the underyling substrate.

Because the erosion rate of low-Z materials via physical and chemical sputtering is larger than for high-Z, such as tungsten, concerns have been raised that high duty cycle devices clad with low-Z PFCs may experience unacceptable levels of tritium retention via co-deposition [3] and/or unsustainable build-up of material "slag" [4]. A rigorous understanding of the sourcing and migration rates of low-Z materials in a fusion environment is thus essential in order to evaluate their potential use in next-step devices.

To this end, this paper examines silicon carbide (SiC), a low-Z, refractory ceramic material, as a PFC for the DIII-D tokamak divertor. Historically, the many attractive properties of SiC motivated studies on various fusion experiments [5][6][7] but perceived technological limitations prevented a closer consideration at the time. The decision to exclude SiC as a plasma-facing material for ITER [8] has resulted in few subsequent evaluations on confinement devices. Progressive advances in the performance of SiC fiber composites ($SiC_f$/SiC) for Gen-IV fission reactors [9], however, have stimulated renewed interest in SiC for fusion applications [10][11][12]. Modern grades of $SiC_f$/SiC share many of the attractive properties of CFCs (thermo-mechanical strength, high melting point, commercially available manufacturing) and may have additional benefits such as:

1) Minimal degradation of themo-mechanical properties even under high levels of neutron damage [13]. This effect becomes particularly pronounced at high temperatures.

2) Substantially reduced chemical sputtering yield relative to graphite [14][15], implying less material slag and lower levels of fuel retention via co-deposition.

3) Lower tritium diffusivity relative to tungsten, which can be further decreased by a factor of ~$10^{-6}$ by depositing a thin (~100 μm) coating of monolithic SiC on top of a SiC/$SiC_f$ substrate – as in the ARIES-AT reactor design study (1 mm SiC on 4 mm $SiC_f$/SiC, [16]).

Fortuitously, siliconization may also be superior to other low-Z wall conditioning techniques like carbonization or boronization in terms of lowering oxygen impurity levels, increasing plasma density limits, and improving energy confinement [17][18].

These inherent advantages, as well as the consensus that the material mix chosen for ITER is unlikely to extrapolate to future machines, motivates a re-evaluation of silicon carbide as a candidate PFC in modern tokamaks under high performance, high heat flux plasma conditions. In this paper some of the first "wind-tunnel" tests of silicon carbide under reactor-relevant heat and particle flux densities are presented with a modern set of PMI diagnostic and model validation capability. Pprevious quantitative SiC erosion studies have been performed only via mass loss on ion beam devices [11][14][19]. These experiments did not distinguish the respective Si and C contributions to the overall erosion rate or the sputtering mechanism (physical vs. chemical). Some attempts have been made using residual gas analysis to infer chemically eroded species [14], but such techniques are difficult to quantify. Similarly, the CD emission rate from SiC at high-flux D plasma bomardment has been monitored previously and compared to graphite [15] but was not absolutely calibrated or quanitfied via the D/XB method. The present work is complemented by a companion paper that compares the erosion and retention behavior of SiC, W, Si, and graphite under D bombardment in linear plasma devices [20].

In Section 2, a silicon carbide erosion model is developed that incorporates erosion via physical and chemical sputtering, as well as surface enrichment/dilution effects due to preferential sputtering and background impurities. Section 3 describes fabrication of the SiC samples used in DIII-D tokamak experiments and the conditions in the under which they were tested. In Section 4 the SiC mixed-material erosion model is validated at divertor-level surface temperatures, ion flux densities, impact angles and energies. Notably the gross erosion of SiC is inferred to be significantly lower than graphite across a wide range of plasma conditions, confirming and extending the results of previous low-power L-mode experiments [12]. Section 5 provides post-mortem surface information on the robustness of the SiC coatings, on a macroscopic and microscopic scale, to morphological and compositional changes under DIII-D divertor heat loading and particle fluence. The paper concludes with a discussion of these results and how they can be extrapolated to next-step devices.

## 2. Silicon Carbide Mixed-Material Erosion Model

The erosion flux of silicon, $\Gamma_{e,Si}$, and carbon, $\Gamma_{e,C}$, from SiC in the DIII-D divertor environment involves bombardment by deuterium main ions with incident flux density $\Gamma_{i,D}$ and C impurity ions with incident flux density $\Gamma_{i,C}$. These processes result in a mixed-material surface consisting of some fraction of elemental C, Si, and covalently bonded SiC. The total erosion yield of Si and C from this mixed-material surface, defined as $Y_{Si,tot} = \Gamma_{e,Si}/\Gamma_{i,D}$ and $Y_{C,tot} = \Gamma_{e,C}/\Gamma_{i,D}$, respectively, can be estimated as

$$Y_{Si,tot} = \sigma_{SiC}Y_{SiC,Si} + \sigma_{Si}Y_{Si} \quad (1)$$
$$Y_{C,tot} = \sigma_{SiC}Y_{SiC,C} + \sigma_C Y_C \quad (2)$$

where $\sigma_{Si}$, $\sigma_C$, and $\sigma_{SiC}$ refer to the fractional abundances of Si, C, and SiC within some characteristic implantation depth $\Delta$ from the plasma-facing surface. In this work it is assumed that no other impurities are present on the surface, i.e, $\sigma_{Si} + \sigma_C + \sigma_{SiC} = 1$. $Y_{Si}$ and $Y_C$ refer to the total sputtering yields of pure Si and pure graphite, and $Y_{SiC,Si}$ and $Y_{SiC,C}$ are the erosion yields of Si and C from SiC, respectively. These erosion yields can be broken into constituent yield terms as follows:

$$Y_{SiC,Si} = Y_{D \to SiC,Si,ph} + Y_{D \to SiC,Si,ch} + f_C Y_{C \to SiC,Si,ph} \quad (3)$$
$$Y_{SiC,C} = Y_{D \to SiC,C,ph} + Y_{D \to SiC,C,ch} + f_C Y_{C \to SiC,C,ph} \quad (4)$$
$$Y_{Si} = Y_{D \to Si,ph} + Y_{D \to Si,ch} + f_C Y_{C \to Si,ph} \quad (5)$$
$$Y_C = Y_{D \to C,ph} + Y_{D \to C,ch} + f_C Y_{C \to C,ph} \quad (6)$$

The subscripts $ph$ and $ch$ differentiate sputtering via either physical or chemical processes. The carbon flux fraction $f_C$ is defined as $\Gamma_{i,C}/\Gamma_{i,D}$ and is assumed to be small, such that $\Gamma_{i,D} + \Gamma_{i,C} \approx \Gamma_{i,D}$. Si impurity content in the plasma and the corresponding physical sputtering (including self-sputtering) of SiC due to Si ion impact is neglected, but the calculated Si erosion rate from SiC is significantly lower than the C source above very small values of $f_C$. Therefore it is expected that $f_{Si} \ll f_C$ and thus ignoring sputtering due to Si impacts is reasonable.

### 2.1 Physical Sputtering of SiC

Physical sputtering (PS) yields of C and Si from Si, SiC, and graphite are calculated using the SDTrimSP Binary Collision Approximation (BCA) code [21] in the static approximation. An ion impact angle of 45 degrees from normal is assumed for all SDTrimSP simulation cases. This is a simplification relative to the spectrum of ion impact angles present in a tokamak divertor [22][23], but since only the total erosion yield can be measured in experiments, this impact angle represents a useful characteristic average value by which to compare expected erosion behavior to spectroscopic measurements (Section 4). It is noted that a previous study in the ASDEX-U divertor calculated an average ion impact angle for D and C closer to 65 degrees [22]. These calculations did not account for finite surface roughness, however, which tends to shift the ion impact angle distribution towards normal incidence (see Fig. 9 in ref. [22]). Because the SiC coatings used in this work had fairly rough surfaces (**Table 2**), an average ion impact angle of 45 degrees is assumed to be more reasonable. The 45 degree impact angle assumption has also been successfully used in other recent PMI model validation studies in DIII-D involving tungsten coatings with similar surface roughness [24][25]. The sensitivity of the calculations to the chosen ion impact angle is discussed in Section 4.2.

Surface binding energies (SBEs) for Si and C are derived using the standard enthalpy of sublimation model. For silicon carbide, SBEs for each crystallographic orientation- (110), (111) C-rich, and (111) Si-rich- are calculated from their respective inter-atomic potentials, which are in turn estimated using an analytic formulation for ceramic materials by Tersoff [26] based on a bond-order approach. This formulation has demonstrated reasonable agreement with short-range molecular dynamics (MD) simulations of SiC physical sputtering [10]. It should be noted that the inferred SBEs (provided in **Table 1**) differ significantly from a simple linear superposition of the SBEs of the constituent elements- essentially an amorphous Si:C mixture- which is the default assumption in SDTrimSP for a binary target material.

| Target Material | Surface Binding Energy (eV) | |
|---|---|---|
| | Silicon | Carbon |
| SiC(110) | 16.62 | 10.51 |
| SiC(111) Si-rich | 14.04 | 11.1 |
| SiC(111) C-rich | 18.21 | 7.21 |
| Elemental Si | 4.66 | -- |
| Elemental C (graphite) | -- | 7.43 |

**Table 1:** Calculated surface binding energies (SBEs) of Si and C for each of the three crystalline planes of SiC used in this work, reproduced from [10]. The standard SBE values for elemental Si and C (graphite) are also provided for reference.

The calculated SiC physical sputtering yields for D projectiles are plotted in **Figure 1** along with comparisons to pure silicon, graphite, and amorphous Si:C (1:1 atomic ratio). Unsurprisingly, the highest sputtering of Si occurs from the (111) Si-rich plane and the highest sputtering of C occurs from the (111) C-rich plane, but PS of SiC is not overly sensitive to crystallographic orientation. The results for C projectiles (not shown) are qualitatively similar but shifted to lower impact energies and higher yields. Note that the energy threshold for Si PS from SiC is significantly higher than pure Si, leading to lower values of $Y_{SiC,Si}$ relative to silicon or amorphous Si:C in the low impact energy regime relevant for divertor plasmas. This difference is due to the higher SBE for Si obtained using the Tersoff bond-order approach [26] relative to the linear superposition approach used for amorphous material. The threshold energy and PS yields of C from SiC are similar to the Si:C amorphous material because the SBE of C in SiC is similar to the SBE of pure graphite (see **Table 1**).

### 2.2 Chemical Sputtering of SiC

Chemical sputtering is a temperature-dependent erosion process resulting in the enhancement (or suppression) of the sputtering yield of certain materials under hydrogenic ion bommbardment due to chemical effects. This effect is particularly pronounced for carbon-based substrates such as graphite [27]. One of the promising features of SiC is the observation of reduced chemical sputtering relative to graphite; studies on ion beams [14] and linear plasma devices [15] indicate that the chemical source of C from SiC is a factor of 5-20× lower than graphite. The ratio of $Y_{D \to C,SiC,ch}$ to $Y_{D \to C,ch}$ is somewhat dependent on ion flux and surface temperature, but data in the available literature are too sparse to draw any definitive trends. Therefore, in this work $Y_{SiC,C,ch}$ is simply set equal to $0.1 Y_{C,ch}$, i.e., the chemical erosion of C from SiC is 10% as large as the erosion from graphite, which is calculated using the standard Roth formula [27].

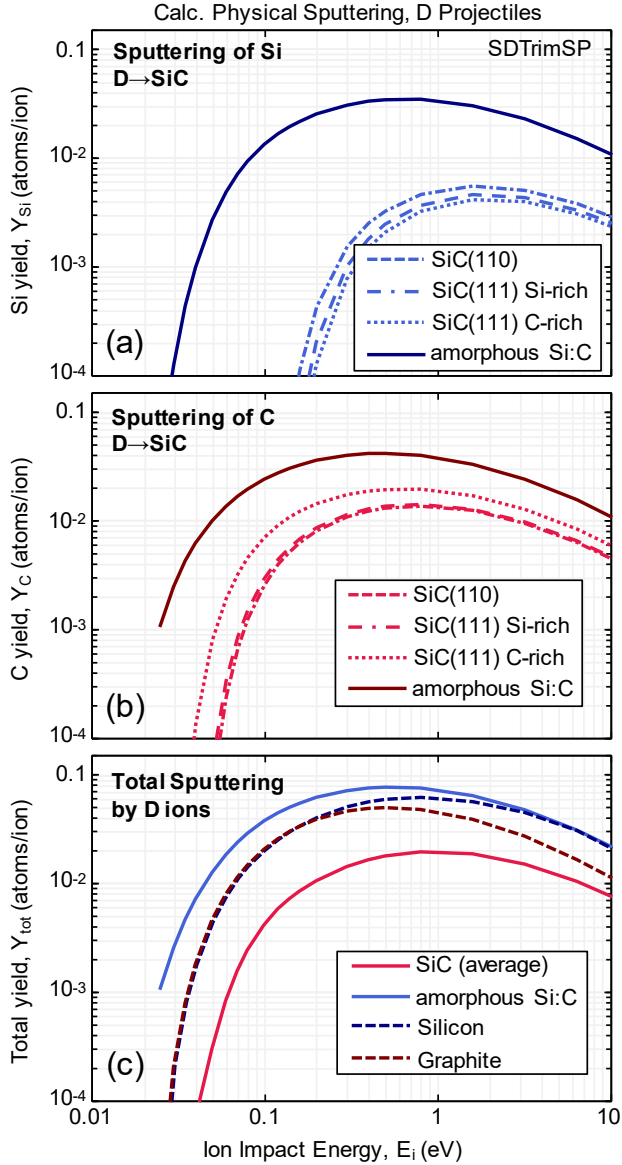

**Figure 1:** Physical sputtering yields of (a) Si and (b) C for D projectiles incident on each crystallographic orientation of SiC, as well as amorphous Si:C, calculated by the SDTrimSP code. (c) Total physical sputtering yields for D projectiles on Si and C materials.

No chemical sputtering of Si from SiC has been observed using mass spectrometry during D ion beam irradiation [14] or spectroscopically on linear plasma devices [20]. It is possible that chemical sputtering of Si from SiC occurs at levels below measurement thresholds, but absent any evidence of this effect, the values of $Y_{D\to SiC,Si,ch}$ are set uniformly to zero. In constrast, chemical sputtering of Si has been routinely observed during D ion irradiation of pure silicon crystals [14][27]. The chemical sputtering of Si is dependent on both surface temperature and ion impact energy, although the dependencies are quite different than for graphite. Values of $Y_{D\to Si,ch}$ used in this work have been interpolated (and in some cases extrapolated) based on the available literature.

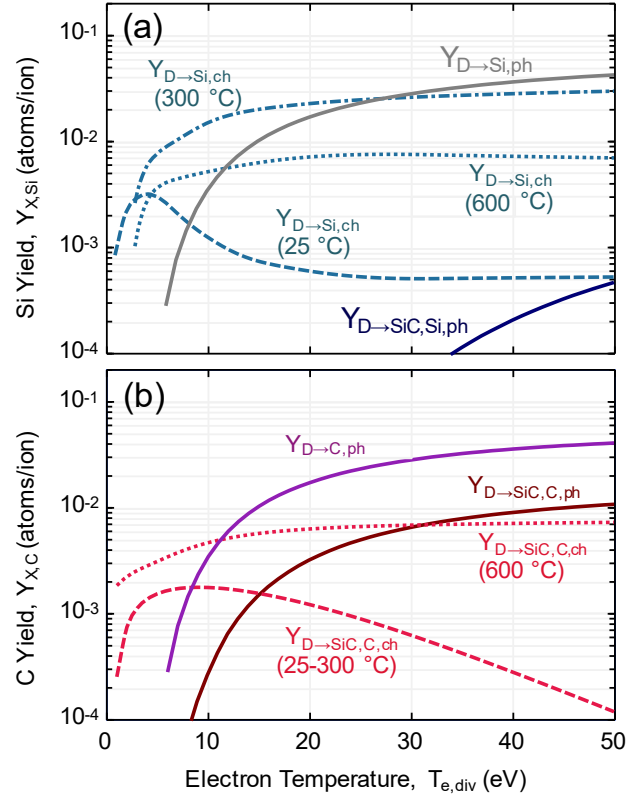

**Figure 2**: Predicted erosion yields of (a) Si and (b) C from Si, C, and SiC due to physical and chemical sputtering as a function of divertor electron temperature, assuming a Maxwellian ion energy distribution with sheath potential of $3T_{e,div}$. For SiC, the average erosion yield from all 3 crystallographic orientations is plotted. Chemical erosion yields are provided for several representative surface temperatures only.

The dependency of these physical and chemical sputtering yields on divertor electron temperature, $T_{e,div}$, is calcuclated assuming a 3D Maxwellian ion impact energy distribution shifted by $3ZT_{e,div}$ to account for the sheath potential drop, where $Z$ is the average charge state of the impacting ions. A value of $Z=1$ is used for D and a value of $Z=2.5$ is assumed for C based on previous OEDGE/DIVIMP modeling [24]. The plasma is assumed to be sufficiently collisional such that the main ion, electron, and impurity ion temperatures are equal.

The inferred physical and chemical sputtering yields for these materials under D ion irradiation are plotted in **Figure 2** as a function of $T_{e,div}$ for several representative surface temperatures. Due to the aforementioned differences in SBE, the physical sputtering yield of C from SiC is reduced by approximately 75% relative to graphite. Due to the low chemical sputtering yield of SiC, the C erosion from SiC is still strongly dominated by physical sputtering except at very low divertor electron temperatures and high surface temperatures. Incredibly, the physical sputtering of Si is reduced by about 99% relative to pure silicon. Under most conditions the Si erosion source from SiC will be primarily due to Si surface enrichment and subsequent erosion of this partially pure Si layer. Chemical erosion of silicon becomes significant at relatively low surface temperatures. For $T_{surf}$ = 300 °C, Si chemical sputtering is comparable or larger than the physical sputtering, regardless of the value of $T_{e,div}$. Si chemical sputtering decreases at higher surface temperatures. The chemical erosion of C also begins to decrease when $T_{surf}$ > 600 °C, as has been characterized in the literature (e.g., see [26] and references therein). Surface temperatures higher than 600 °C are out of the scope of this study, however, because they were not reached in the DIII-D divertor during these experiments.

*2.3 SiC-C-Si Material Mixing*

Following the approach in [28], the regimes of net erosion and net deposition of impurities in a tokamak divertor can be estimated analytically assuming the surface quickly reaches equilibrium. This condition can be expressed as $\left(\frac{1}{\rho}\frac{d\rho}{dt}\right)^{-1} \ll \tau$, where $\rho$ is the areal density of the impurity in the mixed-material surface layer and $\tau$ is the variation time of the background plasma parameters. Here we approximate $\rho$ as $n_Z \Delta$, where $\Delta$ is the depth of the mixed-material layer and $n_Z$ is the atomic density of impurity $Z$ (assumed uniform throughout the layer). Using the ion implantation depth for $\Delta$, $n_{SiC} \sim 10^{29}$ m$^{-3}$, and estimating $d\sigma/dt$ as $Y\Gamma_{i,D}$, we obtain a value of several hundred ms for $\left(\frac{1}{\rho}\frac{d\rho}{dt}\right)^{-1}$ for characteristic DIII-D parameters ($\Delta$ = 10 Å, $\Gamma_{i,D} = 10^{22}$ m$^{-2}$ s$^{-1}$ and $Y$ = 0.03). This is much shorter than the typical DIII-D current flat-top duration (5 s) and thus the eqilibrium assumption is considered reasonable.

In the net C deposition regime, the C coverage will monotonically increase until the entire SiC surface is covered with carbon, i.e., $\sigma_C \equiv \rho_C / \rho_{graphite} = 1$. In a net C erosion regime, a mixed-material fraction of carbon, $\sigma_C < 1$, will be established and can be estimated by the relation

$$\sigma_C = \frac{(1-R)f_C}{Y_C} \qquad (7)$$

where $R$ is the reflection/backscattering coefficient of C ions on the SiC surface (calculated to be ~0.1 via SDTrimSP).

On the portion of the SiC surface that is not covered in carbon the stiochiometric ratio of Si to C will increase because the total erosion yield of C from SiC is higher than Si from SiC for all divertor conditions (**Figure 2**). This preferential sputtering process results in a partially enriched Si layer of surface fraction $\sigma_{Si}$ that is not chemically bound to the SiC substrate. An expression for $\sigma_{Si}$ in terms of $\sigma_C$ and sputtering yield terms can thus be derived by noting that, in equilibrium, $\sigma_{SiC}Y_{SiC,Si} + \sigma_{Si}Y_{Si} = \sigma_{SiC}Y_{SiC,C}$, i.e., the atomic erosion rate of Si from SiC and the partial pure Si layer is equal to the erosion rate of C from the SiC layer. Solving for $\sigma_{Si}$ we obtain:

$$\sigma_{Si} = (1-\sigma_C)\frac{Y_{SiC,C} - Y_{SiC,Si}}{Y_{Si} + Y_{SiC,C} - Y_{SiC,Si}} \qquad (8)$$

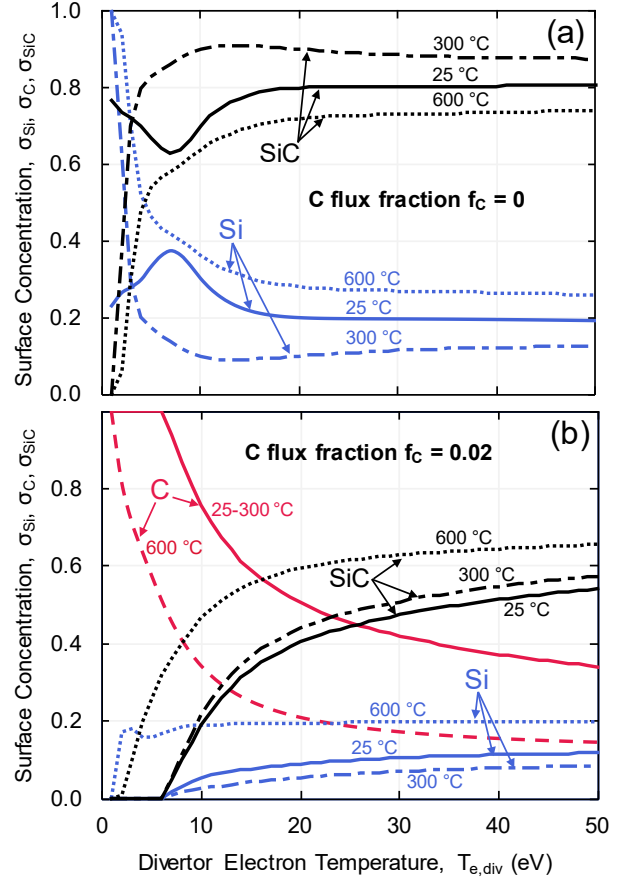

**Figure 3**: Calculated equilibrium surface concentrations of Si, C, and SiC as a function of divertor electron for (a) the special case of no incident C impurity flux (fc=0) and (b) the more typical C flux fraction of 2% (fc=0.02). Calculations for three representative values of surface temperature are provided.

The re-deposition of eroded Si back onto the surface and any corresponding self-sputtering effects are neglected due to the minimal expected impact relative to the erosive C source in DIII-D.

Equilibrium SiC-C-Si mixed-material concentrations are plotted at several different surface temperatures in **Figure 3** for the special case $f_C = 0$ and a typical DIII-D value $f_C = 0.02$. In the absence of C deposition, significant Si surface enrichment occurs at low divertor electron temperatures (detached plasma state) because the threshold for C physical sputtering from SiC is lower than for Si. At higher electron temperatures, where physical sputtering dominates, the erosion yield of C from SiC is only slightly higher than Si from SiC, so the Si enrichment is relatively modest. As previously noted, because the value of $Y_{SiC,C}$ is higher than $Y_{SiC,Si}$ for the entire range of $T_{e,div}$ and $T_{surf}$ examined in **Figure 3**, there is no regime in the $f_C = 0$ case with C enrichment of the SiC surface. The "roll-over" point for Si chemical sputtering yields is at the relatively low surface temperature of ~300 °C, so $\sigma_{Si}$ is minimized here and then increases somewhat from 300 °C to 600 °C.

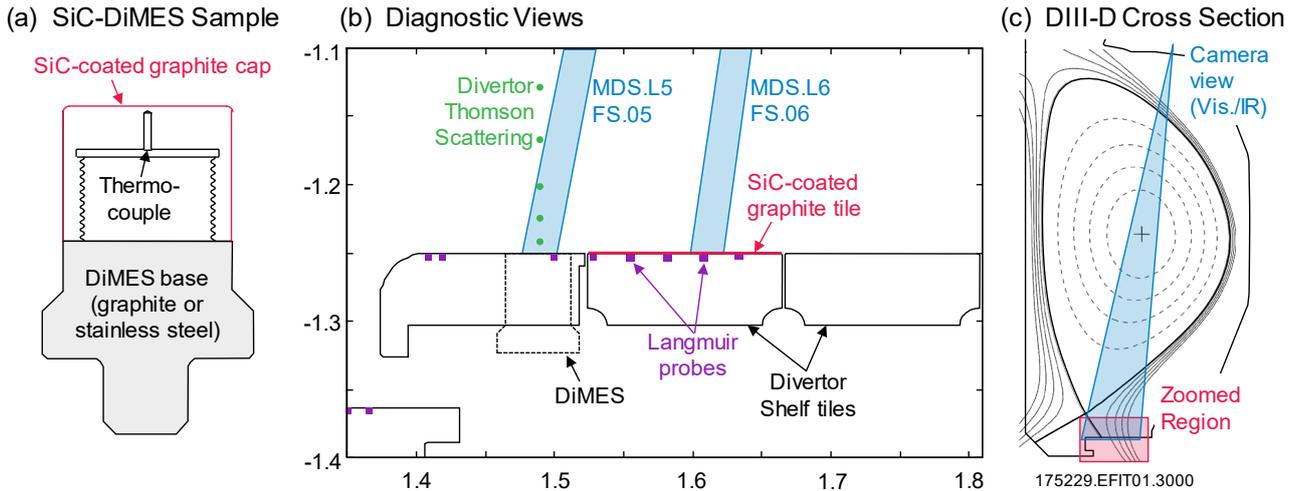

**Figure 4:** (a) Cross-section of the SiC-coated graphite DiMES sample. (b) Zoomed-in diagram of the DIII-D lower divertor shelf region, indicating the diagnostics used in this study, as well as the location of DiMES and the SiC-coated tiles. (c) DIII-D poloidal cross section with typical magnetic equilibrium reconstruction overlaid.

In contrast, Si enrichment is minimal at low electron temperature for the $f_C = 0.02$ case because the SiC surface becomes heavily diluted with carbon. At higher values of $T_{e,div}$ (more attached plasma), the physical and chemical erosion of carbon is more efficient and thus the equilibrium C mixed-material fraction decreases. The purity of the SiC surface increases from 25 °C to 300 °C due to enhanced chemical erosion of the enriched Si layer. From 300 °C to 600 °C, the SiC surface purity further increases due to the large decrease in $\sigma_C$, which dominates over the small increase in $\sigma_{Si}$.

## 3. Experimental Apparatus and Procedure

### 3.1 Sample Preparation

Graphite samples (GrafTech Inc., Grade ATJ™) were fabricated to the standard geometry used by the DIII-D Divertor Materials Evaluation System (DiMES) [29], indicated in **Figure 4a**, and to the standard tile dimensions for the DIII-D lower divertor, shown in **Figure 4b**. Deposition of silicon carbide material was performed in-house at General Atomics. The SiC coating was applied via chemical vapor deposition (CVD), which in this case utilized of the decomposition of methyltrichlorosilane, $CH_3SiCl_{s(g)}$, in a high-temperature hydrogen environment [30]. This method for the fabrication of SiC composite materials for nuclear application results in a high purity, crystalline β-SiC that has previously demonstrated good performance in fission irradiation environments [31][32].

The graphite tiles and DiMES caps were loaded into a high-temperature vacuum furnace with deposition carried out under vacuum and at elevated temperatures. The resulting deposited SiC layers were 100-200 μm thick. Despite best efforts to mask the bottom edge of the DiMES samples from SiC deposition, some SiC material appeared on the bottom edge and inner threads of the caps after the CVD process. This necessitated manual sanding/grinding of the bottom and inner threads of the cap in order to properly fit it on to a stainless steel DiMES base (**Figure 4a**) as to be flush with the DIII-D divertor shelf. Alignment with the surface of the divertor shelf tiles was verified to a tolerance of 50 μm. The samples were only handled with gloves during the alignment process, but some impurities were likely introduced into the SiC coating due to the sample handling, as indicated by *in-situ* visible spectroscopy (Section 4) and post-mortem surface analysis (Section 5).

### 3.2 Experimental Setup

The results presented in this paper represent the compilation of a series of experiments, some dedicated and some "piggyback," carried out during the 2017-19 DIII-D run campaigns. DIII-D is a medium-scale tokamak with major radius R=1.6 m, minor radius a=0.67 m, pulse lengths of ~5 s, on-axis magnetic field of 2.1 T, and typical plasma currents ranging from $I_p$=1-2 MA. The typical magnetic geometry and diagnostic setup for these experiments is shown in **Figure 4c**.

As a first test of the robustness of SiC coatings in the tokamak divertor environment, a piggyback experiment was carried out during the 2017 campaign. A SiC-coated graphite DiMES sample was exposed to 29 consecutive plasma discharges with the outer strike-point (OSP) on the divertor shelf (**Figure 4b**). The experiment consisted of both L-mode and H-mode discharges, with injected power, $P_{INJ}$, ranging from 3-9 MW, steady-state heat fluxes on the SiC target of 1-3 MW/m$^2$, and transient heat fluxes due to ELMs of 5-20 MW/m$^2$. Some discharges were "fixed-OSP" shots in which the OSP was fixed near the inner edge of the DiMES sample ($R_{OSP}$=1.46 m) and others were "swept-OSP" shots where the OSP was moved rapidly between $R_{OSP}$=1.40-1.65 m. In total the sample was exposed to ~150 s of plasma discharges, approximately 50% in L-mode and 50% in H-mode. The cumulative D ion fluence to the sample was approximately $10^{24}$ atoms/m$^2$ as inferred from D spectroscopy diagnostics using the S/XB method and divertor Thomson scattering (DTS) described below. Dedicated follow-up SiC-DiMES experiments were subsequently performed in 2018 and 2019 at similar plasma conditions but with more controlled scans of $T_{e,div}$ and $T_{surf}$.

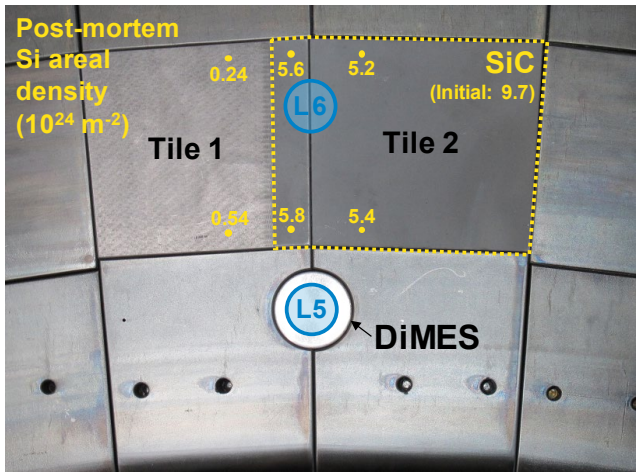

**Figure 5:** Top-down photograph of the SiC-coated graphite tiles installed in the DIII-D lower divertor. The location of DiMES and the MDS L5/L6 chords are labelled. Post-mortem measurements of the Si areal densities at various locations across these tiles (discussed further in Section 5.2) are also provided.

Finally, high-fluence tests were performed on two tiles coated with 100±50 µm of SiC installed in the DIII-D divertor shelf for the entirety of the 2018 run campaign. Both tiles were located just radially outboard of the DiMES port, as depicted in the poloidal cross-section in **Figure 4b**. SiC coatings were deposited on the entire surface of the tile on the high toroidal angle side ("Tile 2") and on the first 2 cm in the toroidal direction on the neighboring tile ("Tile 1"), as shown in **Figure 5**. The SiC coating was milled from remainder of Tile 1 to minimize Si contamination of DiMES via material migration from the SiC-coated tiles. The majority of DIII-D experiments are conducted with a clockwise magnetic field direction, resulting in impurity migration in the high-angle toroidal field direction due to plasma flow, and in the radially inboard direction by E×B drifts (top-left to bottom-right in **Figure 5**). The 2 cm coating width was selected to encompass the viewing spot of the multichordal divertor spectrometer (MDS) L6 chord, also shown in Figure 5. Because the SiC-coated tiles were present for an entire DIII-D campaign they were exposed to a wide variety of divertor plasma conditions with heat fluxes ranging from approximately 0.1-5 MW/m$^2$ in steady state and 5-50 MW/m$^2$ during ELMs. These tiles were exposed to a cumulative D fluence of about $3\times10^{25}$ m$^{-2}$ over ~$1.3\times10^4$ s of plasma operation.

The divertor plasma was characterized primarily by the lowest chord of the DIII-D DTS system [33], located 8 mm above DiMES surface (but at a different toroidal location) and operated at 50 Hz (**Figure 4b**). The DIII-D shelf Langmuir probes were used to verify consistency with DTS measurements. The heat flux to the divertor shelf was measured with ~5 mm radial resolution using infrared thermography of the surrounding graphite divertor tile surfaces. These measurements were also used as a proxy for the time evolution of the SiC surface temperature. Peak temperatures between 200 °C and 600 °C were measured during various DIII-D discharges depending on the exposure duration and heat flux density. Thermocouples embedded several mm below the plasma-facing surface indicated that the bulk temperature generally did not exceed 100 °C and returned almost to room temperature between shots. Recycling and impurity influxes from the plasma-facing surfaces were characterized via multiple divertor spectroscopy diagnostics, also shown in **Figure 4b**. The DIII-D high-resolution multichordal divertor spectrometer [34] with ~0.1 Å wavelength resolution and 5 Hz temporal resolution was used to distinguish Si and C line emission signals from the continuum background and potential contaminant lines. A CCD camera [24] with Si-II (6360 Å) and C-II (5120 Å) bandpass filters provided fast (100 Hz) imaging of the spectral line emission from the divertor surface at ~1 mm/pixel to diagnose unipolar arcing behavior of the SiC surfaces during ELMs. Little to no arcing activity was observed, in contrast to previous studies of thin (0.1-1 µm) high-Z coatings in the DIII-D divertor where significant arcing occurred [35].

## 4. Results

### 4.1 Spectroscopic Analysis of SiC Erosion

The overall (physical + chemical) gross erosion rate of Si and C from the SiC samples was inferred using the ionizations/photon, or S/XB, method [36]. This method involves converting absolutely calibrated spectroscopic intensities into removal rates of atoms from the material surface. This method is typically conducted using a neutral emission line, but no strong neutral lines of either Si or C lie within the detection range of MDS so the strong Si-II doublet at 6347.1/6371.4 Å and the C-II singlet at 4267.0 Å were analyzed instead. The 4267.0 Å C-II line was chosen because it lies in close proximity to the CD spectral emission band near 4300 Å and thus both signals can be observed simulataneously within the ~80 Å spectral width of the MDS grating.

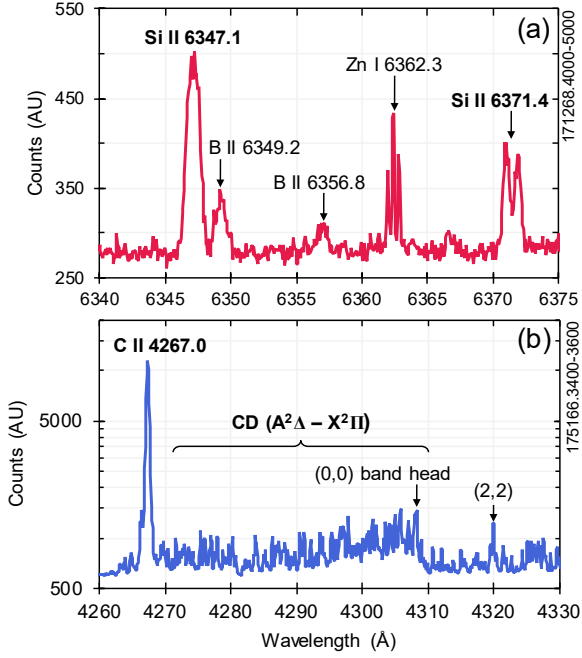

**Figure 6**: Example emission spectrum observed from the DIII-D high-resolution divertor-viewing spectrometer in the region of (a) the 6347.1/6371.4 Å Si-II doublet and (b) the 4267.0 Å C-II line and ~4300 Å CD emission band (note the semi-logarithmic scale).

*4.1.1 Silicon S/XB Analysis*

Si-II emission intensity was converted into absolute flux densities of eroded Si atoms, $\Gamma_{e,Si}$, via the formula

$$\Gamma_{e,Si} = \frac{4\pi}{1-F_{Si}} \left(\frac{S}{XB}\right)_{Si-II} I_{Si-II} \qquad (9)$$

where $\left(\frac{S}{XB}\right)_{Si-II}$ is the ionizations/photon coefficient (cm$^3$ s$^{-1}$) for the Si-II 6360 Å doublet obtained from the ADAS database [38] and $I_{Si-II}$ is the measured spectral emission itensity from the Si-II line (ph cm$^{-2}$ s$^{-1}$ sr$^{-1}$). The dependence of the Si-II S/XB coefficient on divertor electron density and temperature is accounted for; values range from 20-30 at low $T_{e,div}$ and from 15-20 at high $T_{e,div}$. The coefficient $(1-F_{Si})^{-1}$ is a correction factor to account for the fraction of eroded Si atoms that are subsequently ionized and locally re-deposited as Si$^+$ ions (i.e., without ionizing to Si$^{2+}$). The local re-deposition fraction $F_{Si}$ is estimated using an analytic formula for prompt re-deposition [39]. Inferred values of $F_{Si}$ typically range 0.1-0.5 and thus can become a significant correction to the S/XB analysis.

A typical emission spectrum obtained by the MDS system in the Si-II wavelength range is displayed in **Figure 6a**. Several contaminant lines of Zn-I and B-II are present within the Si-II emission region. B-II emission is reguarly observed in DIII-D and is attributed to deposition/re-erosion of boron material introduced via glow-discharge boronization and/or impurity powder dropping [36]. The Zn-I line was unexpected as zinc is not a typical impurity source in the DIII-D tokamak. Careful analysis of the Zn-I line structure (**Figure 7a**, inset), however, confirms exceptionally good agreement with the expected center wavelength and Zeeman splitting pattern of a Zn-I 6362.3 Å line obtained from the NIST database. The evolution of the intensity of these B and Zn lines over the course of 19 consecutive discharges on the SiC-coated DiMES sample is shown in **Figure 7**. With the exception of the first shot in the series, the ratio of Zn-I to Si-II emission decays exponentially. This suggests a trace zinc impurity layer on the SiC surface is gradually eroded away by repeated plasma exposures. This hypothesis is further confirmed by trace amounts of Zn observed on the sample surface during post-mortem analysis (Section 5.2). In contrast, the ratio of the B-II to Si-II emission lines remains relatively constant throughout the sequence, varying between ~0.1-0.3 for fixed-OSP discharges and ~0.4-0.6 for swept-OSP discharges. Sweeping of the strike-point may have the effect of liberating re-deposited B layers from previous shots, leading to transiently higher B contamination.

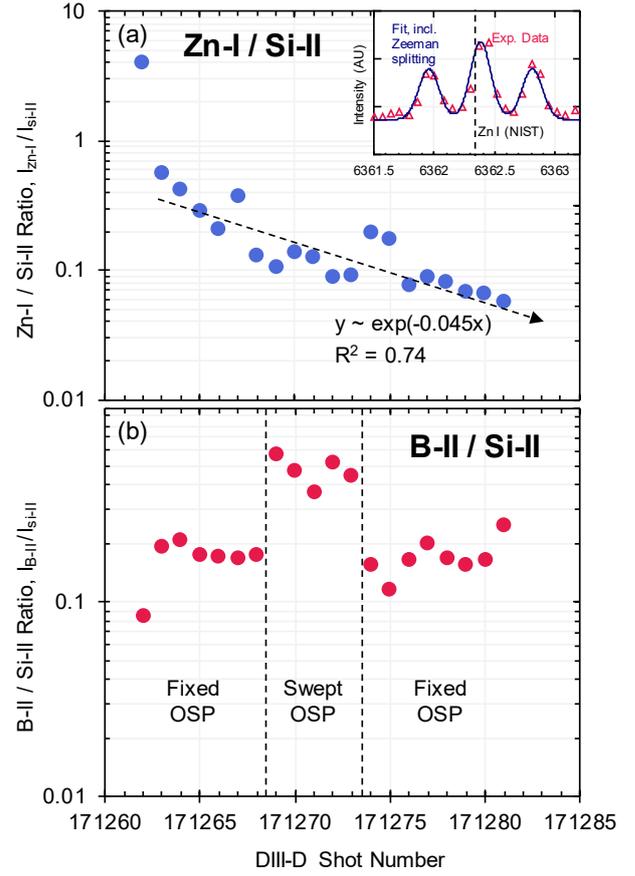

**Figure 7:** (a) Evolution of the ratio of the intensity of the contaminant (a) Zn-I line and (b) B-II lines to the intensity of Si-II doublet over the course of progressive discharges in this SiC-DiMES experiment.

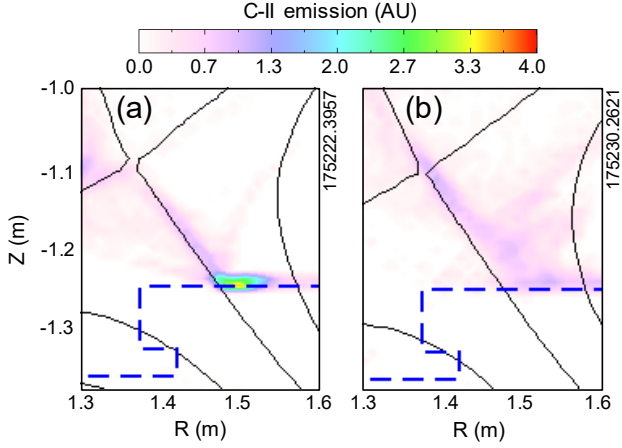

**Figure 8**: Poloidal reconstruction of the C-II emission profile near the DIII-D divertor surface for (a) a high-temperature, attached divertor plasma phase ($T_{e,div}$ = 20 eV) and (b) a low-temperature, detached divertor plasma phase ($T_{e,div}$ = 1 eV).

Fortunately the B-II and Zn-I peaks do not overlap directly with the Si-II peaks, permitting an accurate background subtraction using line-fitting techniques. When interpreting Si-II emission data from bandpass-filtered diagnostics, these contaminanant lines could compromise spectroscopic measurements of Si gross erosion. Such effects have been observed in previous DIII-D studies of tungsten erosion during ELMs where fast, filtered measurements were essential [40]. This is outside the scope of the present work because the primary physics of interest is not intra-ELM effects, but care must be taken to minimize these contaminant impurity sources if and when such measurements are needed. This could be accomplished, for example, by running repeat cleaning discharges with the strike-point fixed in the same position.

*4.1.2 Carbon S/XB and D/XB Analysis*

The identical analysis procedure in Equation (8) is used to infer the C erosion flux density, $\Gamma_{e,C}$, with S/XB coefficients and measured emission intensities replaced with their respective values for the C-II 4267.0 Å spectral line. Values for the C-II S/XB coefficient range from approximately 130 to 170 within the range of $T_{e,div}$ and $n_{e,div}$ considered in this work. A typical emission spectrum obtained by MDS in this wavelength range is shown in **Figure 6b**. The prompt re-deposition factor for C$^+$ is very low in all cases ($F_C$<0.01) and is thus neglected. The chemical component of the C gross erosion rate from SiC, $\Gamma_{e,C,ch}$, is also measured using the molecular CD (A$^2\Delta$ – X$^2\Pi$) transition for methane break-up located near 4300 Å [41]. The following H-mode scaling derived from ASDEX-U measurements [42] is used to calculate the corresponding disassociations/photon coefficients (D/XB):

$$\left(\frac{D}{XB}\right)_{CD} = -55 + 31.9 \cdot n_{e,19}^{0.2} + 24.2 \cdot T_e^{0.1} \quad (10)$$

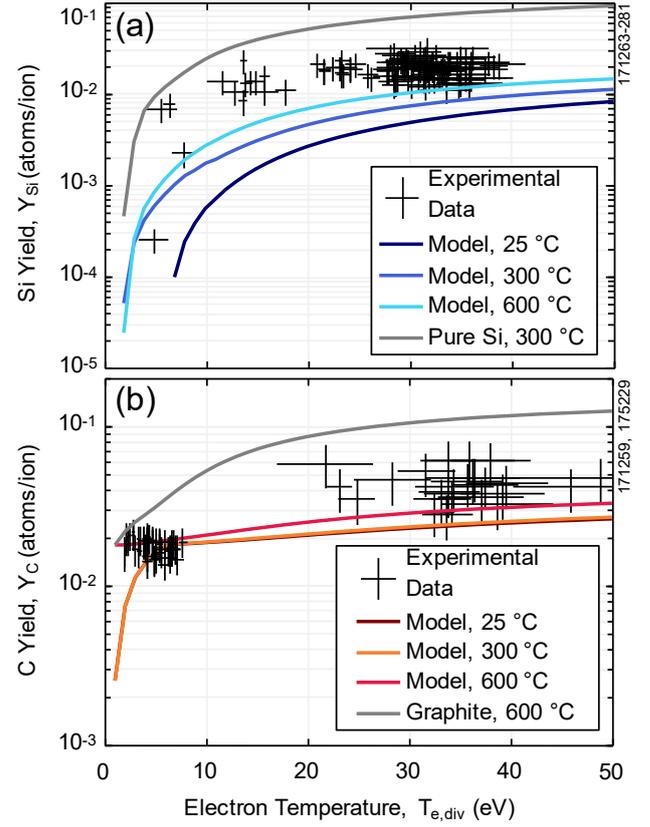

**Figure 9:** Measured total erosion yields of (a) Si and (b) C during exposures of SiC-coated DiMES samples in DIII-D. Predictions from the SiC erosion model described in Section 2 are overlaid. Also included are the corresponding total erosion yields from pure Si and pure graphite at the approximate $T_{surf}$ value corresponding to peak chemical sputtering.

where $n_{e,19}$ is the divertor electron density in units of $10^{19}$ m$^{-3}$ and $T_e$ is the electron temperature in eV. Values of the D/XB coefficient used in this work range from ~40 at $T_{e,div}$ = 2 eV down to ~20 at $T_{e,div}$ = 15 eV.

Due to the relatively high C-II background level in DIII-D, C erosion yields from the SiC-DiMES surface can only be reliably inferred via the C-II line in strongly attached conditions where C-II radiation is localized strongly near the divertor surface. Poloidally-resolved C-II radiation patterns from the 5140 Å emission cluster, obtained from the DIII-D tangential camera imaging diagnostic [43], are shown in **Figure 8** for the case of $T_{e,div}$ = 20 eV and $T_{e,div}$ = 1 eV. In the high- temperature case the C-II radiation front is very close to the surface, indicating that nearly all C-II radiation originates from material sputtered directly from the divertor target. In the low-temperature case, however, C-II emission is distributed over a broad portion of the divertor leg, indicating significant signal contamination from non-local transport of C$^+$ ions.

Carbon erosion at very low values of $T_{e,div}$, however, can be assumed to be due completely to chemical sputtering and thus the D/XB method provides a reliable estimate of $Y_C$. No reliable measurements could be obtained in the intermediate range (8 eV < $T_{e,div}$ < 20 eV) where C physical sputtering becomes important but C-II emission is not sufficiently localized near the divertor plate. At temperatures where D recombination becomes significant relative to electron-impact ionization ($T_{e,div}$ < 1.8 eV) measurements must also be also discarded because the atomic physics assumptions implicit in the D/XB method are no longer valid. Values of C chemical erosion from SiC could be inferred at very low values of $T_{e,div}$ using a more complex edge physics model, such as the OEDGE suite [44], that incorporates additional volumetric sources and sinks (e.g., recombination, charge exchange, and molecular disassociation/re-association) but such efforts are outside the scope of this paper.

*4.2 Total Gross Erosion of SiC*

Effective erosion yields of Si and C from the SiC surface under plasma exposure were inferred by normalizing the C and Si gross erosion rates measured via the S/XB method (Section 4.1) to the deuterium ion flux to the surface, $\Gamma_{e,Si}/\Gamma_{i,D}$ and $\Gamma_{e,C}/\Gamma_{i,D}$. The D ion flux $\Gamma_{i,D}$ is assumed equal to $0.5 n_{e,div} (2 T_{e,div}/m_D)^{1/2} \sin(\theta)$, where $m_D$ is the D ion mass and $\theta$ is the magnetic incidence angle on the divertor target obtained from magnetic reconstructions (EFIT01).

*4.2.1 Si Erosion*

The measured Si erosion yields from the SiC-coated DiMES sample are plotted as a function of divertor electron temperature in **Figure 9a**. Overlaid are the SiC erosion calculations from the mixed-material model in Section 2 for several representative surface temperatures. Most of the discharges in this experimental series were conducted in strongly attached conditions so most of the available measurements lie in the range $T_{e,div}$ > 20 eV. Insufficient signal-to-noise ratios (SNRs) for $T_{e,div}$ < 10 eV due to low Si-II intensitiy levels hindered additional data analysis. Values of $T_{surf}$ ranging from 100 °C to 600 °C were observed in this experiment but no attempt is made to discriminate between them for the purposes of this study.

Generally it is clear that the measured Si erosion yields lie somewhere between the predictions of the mixed-material model and the calculated sputtering yield curve from a pure Si surface (also overlaid). Once source of this disrepancy may be the effect of edge-localized models (ELMs), which are not included in the mixed-material model described in Section 2. ELMs have the effect of transiently increasing the ion impact energy in the divertor (e.g., see [45]), which causes more Si enrichment of the SiC surface (**Figure 3**) and more Si physical sputtering (**Figure 1**). The difference between measurement and model diverges at low values of $T_{e,div}$, where intra-ELM impurity source begins to dominate over inter-ELM sources [46]. Si physical sputtering yields from SiC are also rather sensitive to the assumed average ion impact angle. For example, adjusting the impact angle for 300 eV D ions to 65° in SDTrimSP results in a 100% increase in the calculated value of $Y_{D \to SiC,Si,ph}$ relative to the default 45° assumption. This suggestions that the average ion impact angle in these experiments may have been slightly steeper than 45 degrees.

As expected from the mixed-material model, the Si erosion yields are significantly higher than the physical sputtering yield of Si from pure SiC calculated in **Figure 2**. $Y_{D \to SiC,Si,ph}$ was calculated to be below $10^{-3}$ even for the highest points in the divertor temperature scan, $T_{e,div}$ ~ 40 eV. The elevated Si erosion yield is not likely explained by chemical sputtering from SiC because there was no spectroscopic evidence of SiD emission during the experiment, consistent with the discussion in Section 2. It is concluded that the dominant source of Si from SiC is siliconization of the surface via preferential sputtering, and the subsequent physical and chemical erosion of this partially-enriched Si layer.

*4.2.2 C Erosion*

The measured C erosion yields from a SiC-coated DiMES sample are plotted vs. $T_{e,div}$ in **Figure 9b**. For the range where C physical sputtering can be assumed to be negligible ($T_{e,div}$ < 8 eV in **Figure 2**) the total C erosion yields are inferred using only the chemical sputtering component measured from the 4300 Å CD emission band via the D/XB method, as detailed in Section 4.1.

At high electron temperatures where physical sputtering of C dominates, the measurement-model comparison looks similar to the Si erosion case. The inferred C erosion yields from SiC lie between the mixed-material model and the expected yield from pure graphite at $T_{surf}$ = 600 °C (calculated via the Roth formula [27]). Again, this may be due to the effect of ELMs, which cause transient increases in the divertor ion impact energy and surface temperature, somewhat enhancing C physical and chemical sputtering, respectively. In addition, if the average D ion impact angle was slightly steeper than the 45-degree value assumed in the SDTrimSP calculations, the physical sputtering yield of C from the SiC surface would increase, resulting in better agreement between modeling and experiment.

In contrast to the measured Si erosion yields, values of $Y_{C,meas}$ do not show a pronounced decrease at low divertor electron temperatures. This is interpreted as chemical erosion of the C surface layer that is deposited on top of the SiC-DiMES sample at low $T_{e,div}$. As shown in **Figure 3**, even for a surface temperature of 600 °C, at least 50% of the SiC-DiMES surface is expected to be covered in deposited carbon for $T_{e,div}$ < 8 eV. Unfortunately this level of C coverage makes it difficult to discriminate between chemical sputtering of pure graphite and silicon carbide in DIII-D. This comparison is shown in more detail in **Figure 10**, which shows the measured C chemical erosion yield from a graphite DiMES sample and a SiC-coated sample as a function of $T_{e,div}$. The measured erosion yields from graphite and SiC are, in fact, nearly identical because, as predicted by the mixed-material SiC erosion model, the SiC surface is mostly coated with C in this electron temperature range. Some small deviation of $Y_{C,ch}$ and $Y_{C,SiC,ch}$ may occur as $T_{e,div}$ increases above 15 eV, but there is too much scatter in the data to make any definitive conclusions.

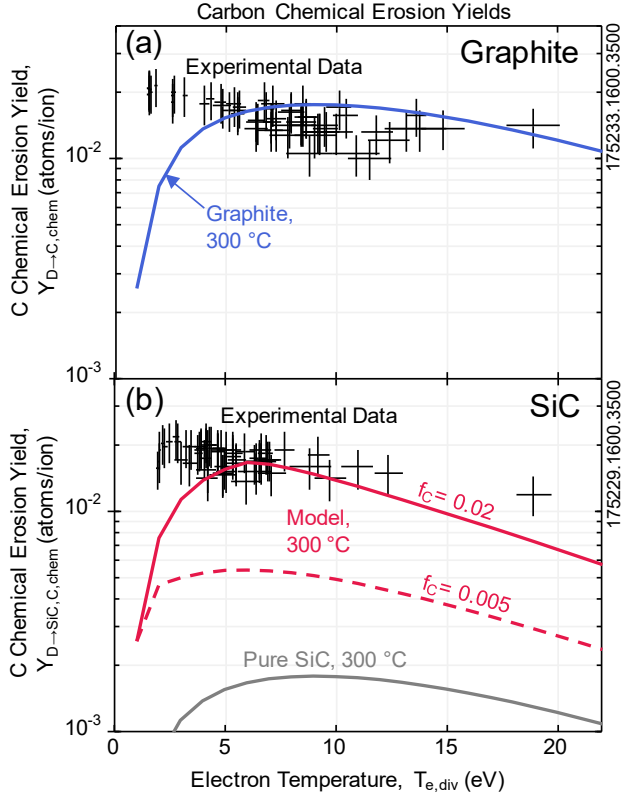

**Figure 10:** Measured C chemical erosion yields from (a) a graphite DiMES sample and (b) a SiC-coated DiMES sample as a function of divertor electron temperature. For graphite, the predictions from the Roth scaling [25] are overlaid. For SiC, the predictions from the mixed-material erosion model are overlaid, along with the expected value for pure SiC (10% of the Roth scaling [15]).

A lower C impurity background level would be necessary to distinguish the chemical erosion yields of graphite and SiC in the DIII-D divertor. As shown in **Figure 10b**, a reduction of the C impurity flux fraction to approximately 0.5% would produce an unambigious test of the C chemical sputtering model for SiC. At $f_C = 0.005$, the peak chemical erosion yield of C from SiC is predicted to decrease by about 70%, while the chemical erosion yield of graphite will not change. Unfortunately such low edge C impurity content is not easily achieved in DIII-D. This could potentially be accomplished by exposing SiC coatings to helium main ion plasmas with a minority D impurity, but to date no such experiments have been performed. He plasma irradiation may also complicate the physics interpretation due to the higher physical sputtering yield and lower energy threshold for C sputtering by He ions.

| Roughness Measurement | SiC-DiMES | | SiC Tiles | |
|---|---|---|---|---|
| | Before | After | Before | After |
| $R_{arith}$ (μm) | 9.4 ± 2.8 | 9.8 ± 2.3 | 5.4 ± 1.6 | 6.2 ± 1.4 |
| $R_{RMS}$ (μm) | 11.9 ± 2.6 | 12.5 ± 2.1 | 6.6 ± 1.5 | 8.0 ± 1.4 |
| $R_{peak-to-valley}$ (μm) | 81.1 ± 9.2 | 107.7 ± 8.1 | 23.2 ± 2.6 | 39.7 ± 3.0 |

**Table 2:** Summary of surface roughness measurements performed on the SiC-coated PFCs before and after exposure to DIII-D divertor plasmas.

A discrepancy is noted between the predicted and inferred C chemical sputtering yields from graphite and SiC at very low divertor electron temperatures ($T_{e,div} < 3$ eV) in **Figure 10**. This is likely due to inaccuracies in the scaling used for the CD 4300 Å D/XB coefficient. The D/XB scaling in [42] was derived only for attached plasma conditions. Subsequent measurements [47] and modeling [48] have indicated the D/XB value may be a factor of 2-3 lower than predicted by the empirical scaling at very low values of $T_{e,div}$. C chemical erosion yields for $T_{e,div} = 1$ eV have been directly measured in DIII-D using the DiMES porous plug injector (PPI) [49] and found to be approximately $5 \times 10^{-3}$, in line with expectations from the Roth scaling [27]. Because the same deviation is present for graphite and SiC, however, the overall conclusion remains that the SiC sample has very similar erosion properties to graphite at low $T_{e,div}$. A lower C impurity content in the DIII-D divertor is required to demonstrate the benefical effect of supressed chemical sputtering of SiC.

## 5. Post-Mortem Analysis of SiC Coatings

### 5.1 Morphological Evolution

No pronounced differences in the morphology of the SiC coatings, e.g., macroscopic delamination or surface cracking, were evident upon visible inspection. Some streaks of carbon deposition were visible, presumably due to migration of C dust from the otherwise full-carbon environment of DIII-D. Profilometry performed before and after the experiment using a WYKO optical interferometer also indicates minimal SiC morphology evolution occurred. A summary of the results of this analysis is provided in Table 2. These roughness values represent the mean of five sampling positions covering the extreme radial and toroidal locations on the DiMES cap, as well as the sample center. The uncertainty was taken to be equal to the standard deviation from the five sampling locations. No change in the root-mean-square (RMS) and arithmetic mean surface roughness were detected within the error bars and only a slight increase in the maximum peak-to-valley distance was observed. The lack of prominent alterations of the SiC surface morphology is consistent with the properties of other refractory ceramics (such as graphite) on which leading edges tend to ablate rather than grow and thus progressive roughening does not occur.

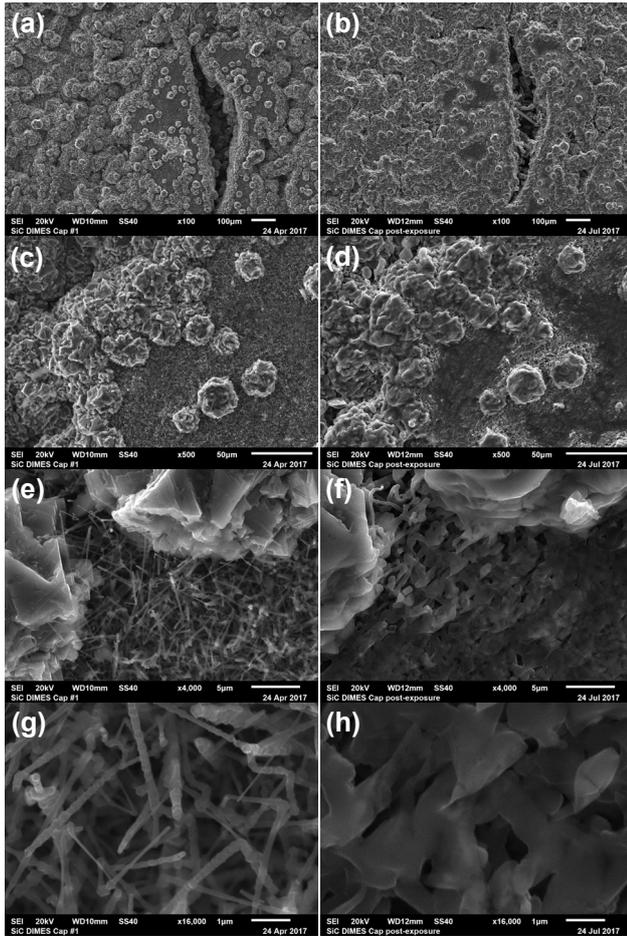

**Figure 11:** SEM micrograph images of the SiC-DiMES sample surface before (a, c, e, g) and after (b, d, f, h) plasma exposure at progressively higher magnification levels (top-to-bottom).

Microscopic analysis of the SiC-DiMES layer was performed via Scanning Electron Microscopy (SEM) on several identical surface locations before and after exposure. Several example images are provided in **Figure 11**. Before exposure, a prominent nodular-like structure with characteristic features 10-50 µm in size is evident, with small SiC "whiskers" ~0.2-0.3 µm in diameter populating the valleys between nodules; this morphology is very characteristic of film growth via the CVD process described in Section 2.1. One large surface flaw approximately 500×100 µm in size (**Figure 11a-b**) demonstrated no significant change due to plasma bombardment, or even perhaps a small decrease in size. Some pillar-like growth is observed inside the flaw, potentially due to deposition/trapping of SiC or C dust grains from elsewhere on the sample. Generally no additional flaking, cracking or surface flaws were observed in the SEM analysis after exposure. Some smoothing/rounding of the nodular particle facets is evident but generally microstructure features of size ≥10 µm remain largely intact. The small SiC whiskers did become "smeared" after plasma exposure, likely due to evaporation/condensation, surface melting, and/or viscous flow processes.

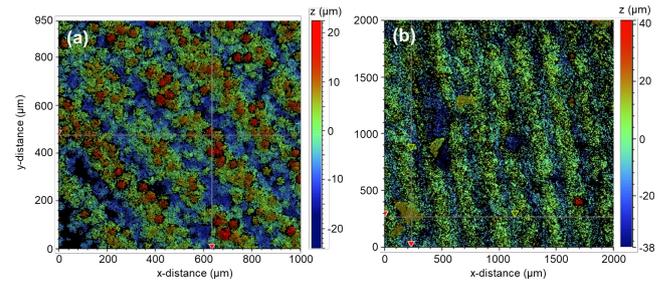

**Figure 12**: False-color image of the surface topology of the SiC-coated surface at the center of Tile 2 (a) before and (b) after a campaign-integrated plasma exposure in the DIII-D divertor.

Similarly, after a campaign-integrated exposure the SiC-coated divertor tiles showed no macroscopic flaking, cracking or other damage upon visual inspection. SEM analysis of the tiles was not possible due to their large size. Profilometry surveys revealed minimal alterations in surface morphology after plasma bombardment. 2D false-color profiles of the surface topography at the center of Tile 2 before and after the campaign are shown in **Figure 12**. The same nodular structures can be observed before exposure due to the CVD process. Periodic ripples appear on the sample surface as artifacts of the tile machining. In contrast to the DiMES cap, the CVD nodules completely dissapear after plasma exposure, indicating that the smoothing/rounding of SiC particle facets further progresses as plasma fluence is increased. Surface roughness values at the center of Tile 2 (indicated in **Figure 5**) measured before and after campaign-integrated exposure are listed in **Table 2**. Due to time constraints, measurements were only performed in one location so error bars are extrapolated by assuming the relative variation in roughness of the surface of the tiles was similar to the DiMES sample. The SiC surface of the tiles was initially smoother than the DiMES cap due to optimizations in the SiC-on-graphite CVD process. Only a small increase in the maximum peak-to-valley distance is observed, likely due to deposition of C or Si dust grains larger than the characteristic surface roughness of the virgin tiles. As with the SiC-DiMES sample, increases in arithmetic and RMS roughness are within the error bars.

### 5.2 Compositional Analysis

Pre- and post-mortem surface composition analysis was performed on the medium-fluence SiC-DiMES sample via energy-dispersive x-ray spectroscopy (EDS). The EDS instrument bombarded the SiC surface with electrons of initial energy 20 keV, resulting in a stopping depth in the material of approximately 3.5 μm. As this depth is much shorter than the thickness of the SiC coating (200±50 μm), it can be assumed that the measured EDS spectrum represents only the SiC layer and not the underlying graphite substrate. Relative atomic concentrations of carbon, $c_C$, and silicon, $c_{Si}$, in this near-surface layer were obtained at different locations across the SiC surface as follows:

$$c_c = \frac{x}{1+x}, \quad c_{Si} = \frac{1}{1+x}, \quad x = f\frac{\hat{c}_C}{\hat{c}_{Si}} \quad (11)$$

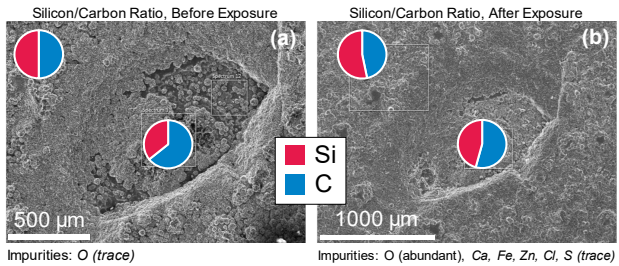

**Figure 14:** Measured Si/C ratio on the surface of the SiC-DiMES sample coating and within a small defect in the coating, (a) before and (b) after plasma exposure, obtained via energy-dispersive x-ray spectroscopy (EDS).

Here $\hat{c}_C$ and $\hat{c}_{Si}$ are the (uncalibrated) atomic concentrations of C and Si, respectively, obtained from the EDS spectra and $f$ is a numerical calibration factor derived from a pristine SiC specimen assumed to contain an exact 1:1 stoichiometric ratio of Si and C.

The carbon/silicon ratios were measured before and after exposure at two identical positions on the sample surface; the results are provided in **Figure 13**. One position represents the standard fully-coated SiC surface that begin with an even stiochiometric ratio (50% Si, 50% C). The other position contains a small defect in the coating on which the graphite substrate was partially visible and correspondingly for which a clear C enrichment is apparent (35.8% Si, 64.2% C). Before exposure only trace oxygen impurities are detected, potentially due to formation of a native oxide layer (SiO$_2$) with a small amount of unbound silicon atoms. After plasma exposure a clear enrichment of surface silicon is apparent on both the exposed SiC surface (53.2% Si, 46.8% C) and inside the defect (45.7% Si, 54.3% Si). The final plasma exposures on this SiC sample were conducted in strongly attached plasma conditions ($T_{e,div} \approx 30$ eV) at surface temperatures close to 600 °C, a regime of net Si enrichment according to the mixed-material model (**Figure 3**). Therefore the increased Si concetration measured by EDS is expected. Si enrichment within the defect may have been due to Si erosion/re-deposition processes from the Si-enriched SiC surface, in conjunction with "shadowing" from re-erosion via ion bombardment. Such effects have previously been observed to be significant in tokamak divertor plasmas [22].

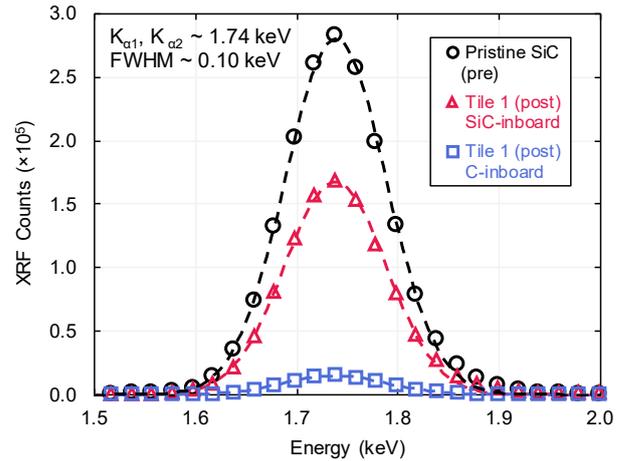

**Figure 13**: Example XRF spectra containing the K$_{\alpha 1}$/K$_{\alpha 2}$ peaks of Si at ~1.74 eV obtained from the inboard side of the SiC-coated and uncoated (graphite) portions of Tile 1, before and after campaign-integrated plasma exposure. Gaussian fits used to obtain the area under the curve are also overlaid.

Significant oxygen contamination of the sample was also evident after plasma exposure. Oxidation likely occurred after sample removal and exposure to atmosphere, causing SiO$_2$ formation at sites from which C atoms were preferentially displaced by kinetic and/or thermal processes. Trace levels of heavier impurities (Zn, Cl, Ca, Fe, S) were also observed. There are small sources of all of these elements inside the DIII-D vacuum vessel, but as discussed above these impurities were more likely introduced during sample handling with nitrile and latex gloves. Zinc, in particular, has been identified as the primary contaminant present in organic samples handled with rubber gloves [50]. The glove manufacturing processes involves catalysts such as zinc oxides and zinc carbonates to enhance the strength and elasticity of the material [51].

The SiC-coated tiles were too large to be analyzed for compositional changes in the EDS chamber. Instead measurements were performed before and after exposure using a hand-held Olympus Innov-X DP-6000 X-ray fluorescence (XRF) analyzer with a beam diameter of about 5 mm. The probing depth of the instrument is on the order of several mm, thus the measured intensity of the silicon XRF signal is proportional to the total areal density of Si on the tile surfaces. Example XRF spectra containing the K$_{\alpha 1}$, K$_{\alpha 2}$ peaks of Si at ~1.74 eV for regions of interest (ROIs) on the SiC-coated and uncoated portions of Tile 1 (taken 10 mm outboard from the inner tile edge) are shown in **Figure 14**. Inferred Si areal densities for all ROIs analyzed are provided in **Figure 5**.

Approximately 40% of the SiC layer (40±20 μm) was removed from the tiles via plasma bombardment and Si deposits ranging from ~1-3×10$^{23}$ m$^{-2}$ in areal density appeared on the uncoated portions of the tiles, presumably due to erosion, migration, and re-deposition processes. Note that since the XRF instrument simply integrates total Si content, no discrimination was available between a pure Si layer on a graphite substrate vs. mixed-material C/Si layer of gradually decreasing Si composition with increasing depth. As expected, more Si deposition is observed on the inboard size of Tile 1 (closer to the outer strike-point) because (a) Si ionization lengths are shorter for hotter, denser divertor plasma, resulting in higher local re-deposition fractions, and (b) as discussed above, the E×B drift direction is radially inboard for the standard DIII-D magnetic field direction. Interestingly, very little radial dependence is observed in the campaign-integrated erosion of Si from the tiles despite typical radial gradients of plasma parameters on the order of 2-3 cm in the divertor region. This suggests that material erosion from the SiC-coated tiles was primarily due to ELMs, which span a much larger plasma-wetted area relative to the inter-ELM phase [52][53].

## 6. Discussion and Conclusions

### 6.1 Silicon Carbide PFCs in Next-Step Devices

We return to a key question investigated in this paper: *does silicon carbide represent a compelling option for next-step devices from the perspective of material erosion and impurity sourcing?*

The model described in Section 2 enables a simple 0D comparison of the total sputter-erosion source (physical plus chemical) between SiC and graphite, the most widely studied and commonly used low-Z PFC. Encouraged by the validation of the mixed-materia SiC erosion model for the wide range of DIII-D divertor conditions described in Section 4, as well as the robust material performance discussed in Section 5, we now extrapolate to reactor-relevant impact energies, surface temperatures, and core plasma conditions. We consider two edge electron temperature regimes: a low temperature, $T_e = 5$ eV, characteristic of "partially detached" divertor plasma expected near the strike-points, as well as low-temperature plasma impacts on main chamber limiter surfaces in the far-SOL. A higher temperature regime, $T_e = 20$ eV, is expected in a "partially detached" divertor scenario away from the strike-points. This is the working physics assumption for the ITER divertor [54]. This edge temperature is also a reasonable proxy for the charge-exchange neutral energy distribution in a DEMO-level device, which is expected to average several hundred eV [55].

For this comparison we assume a uniform PFC temperature of 600 °C. While the equilibrium PFC surface temperature in a reactor may be higher than this, particularly near the strike-points, the empirical scaling for C chemical sputtering in [27] is not considered reliable at higher temperatures due to a lack of experimental measurements on which to constrain it. Furthermore, these calculations are primarily for the purposes of qualitative comparison so conclusions are not expected to be overly sensitive to the specified value of $T_{surf}$. As discussed in Section 2, the silicon impurity concentration in the edge plasma and corresponding physical sputtering via Si ion impact is neglected. We also note that the C impurity fraction in the SOL is obviously correlated with the total C source from the walls rather than being an independetly adjustable parameter. A more "self-consistent" treatment could involve the use of a 2D SOL impurity transport code such as DIVIMP [44] but is beyond the scope of this paper.

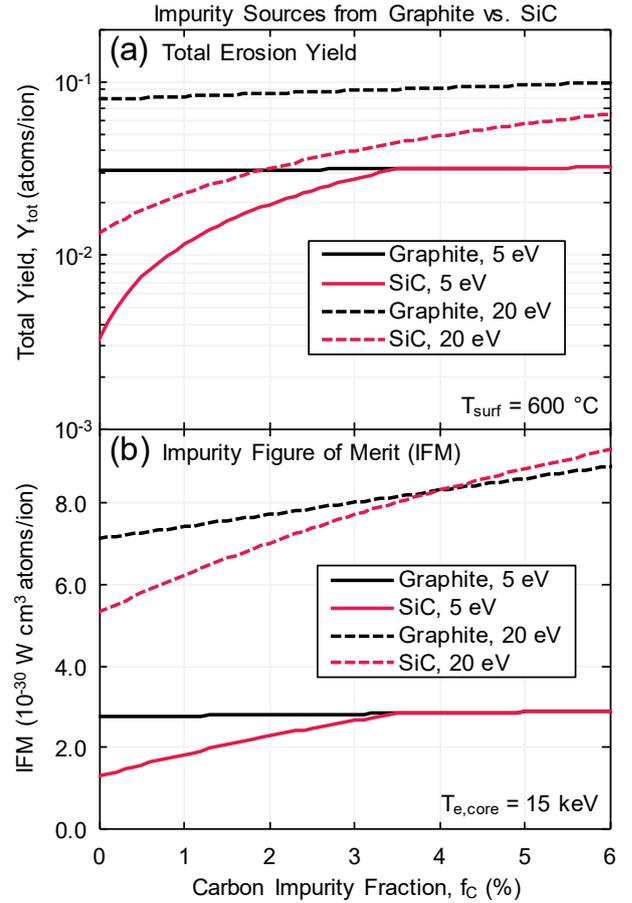

**Figure 15:** Extrapolated (a) total (physical + chemical) erosion yield from all sources, (b) impurity figure of merit, for graphite and SiC PFCs at reactor-relevant conditions, plotted as a function of the carbon impurity fraction at two representative values of edge electron temperature.

The calculated total erosion yield for graphite, $Y_g$, vs. SiC, $Y_{SiC}$, is plotted in **Figure 15a** as a function of the C impurity flux fraction for these two edge electron temperature regimes. In the $T_e = 5$ eV case the total impurity source from SiC is significantly lower than graphite at low values of $f_C$, but converges to the graphite value near $f_C = 0.03$ becuse the SiC surface becomes completely coated in carbon ($\sigma_C = 1$). At $T_e = 20$ eV, the C coverage of SiC is relatively modest and the total erosion source from SiC remains substantially lower than that of graphite for the entire plotted range of $f_C$. C impurity fractions higher than 6% are difficult to imagine, as above this value of $f_C$ the required value of $nT\tau_E$ for ignition begins to increase significantly [4]. These results are encouraging from a wall lifetime perspective. The atomic densities of C and SiC are relatively similar, so the total loss rate of wall material (in mm/year) will be lower for SiC in proportion to the reduction of the erosion yield. This result also has favorable implications for tritium co-deposition rates. Assuming that C and Si co-deposits have similar T trapping probabilities [56], the total T rentention in a SiC wall would be reduced by the factor $Y_{SiC}/Y_g$ relative to a graphite wall.

Finally, in order to evaluate the impact of erosive sources on plasma performance, we introduce an impurity figure of merit (IFM): $P_Z(T_{e,core})Y_Z$, where $P_Z$ is the zero-density radiated power rate coefficient obtained from ADAS [38]. The IFM for a given material is summed over all elemental impurity sources $Y_Z$. A similar metric has been developed for other works [4][57], but has been defined here such that a lower value of the IFM is clearly favorable. A core electron temperature, $T_{e,core}$, of 15 keV has been assumed for comparison, but $P_C$ and $P_{Si}$ are relatively flat in the range of 5 keV to 20 keV, so the exact value of $T_{e,core}$ is unimportant. IFMs for SiC and graphite are plotted as a function of $f_C$ in **Figure 15b**. Although each eroded Si atom radiates about 8× more power than an eroded C atom, a large parameter space exists in which a significant IFM reduction can be achieved with SiC walls relative to graphite. SiC PFCs are particularly beneficial at low C impurity fractions and low electron temperatures- exactly the conditions expected near the strike-points of reactor-level devices. For $T_e = 5$ eV and $f_C = 0$, the IFM for SiC is nearly half that of graphite. At $T_e = 20$ eV, the effect of SiC is less pronouced but a large parameter space still exists ($f_C < 4\%$) where SiC is beneficial.

*6.2 Summary and Conclusions*

In this paper the first "wind-tunnel" tests of silicon carbide as a plasma-facing material in diverted DIII-D H-mode discharges were performed. The principal goal was a first-order evaluation of the erosion properties of the SiC coating. Measurements of the Si and C erosion yields from the material, inferred spectroscopically via the S/XB and D/XB methods, were bechmarked against an analytic mixed-material erosion model. These results represent significant advances in PMI physics understanding over previous erosion studies of SiC on laboratory-scale devices [11][14][19].

A secondary goal of this study was to examine potential surface enrichment effects in SiC at high plasma fluence. This effect was predicted by the mixed-material model and has been observed in some previous studies [58] but not in others [14]. The qualitative trend predicted by modeling- namely, that Si enrichment does occur, and increases with divertor temperature- was confirmed by the present experimental data, both directly through post-mortem compositional analysis and indirectly via *in-situ* visible spectrosocpy. Unfortunately the predicted decrease in the C chemical erosion yield of SiC relative to graphite could not be measured due to the significant C impurity background in DIII-D, which caused the SiC surfaces to become coated with C at low divertor electron temperatures. Future DIII-D studies could be performed in hydrogen or mixed helium/deuterium plasmas where the C impurity content is reduced in order to properly compare the chemical erosion rate of SiC and graphite.

Extrapolating to a DEMO-type device, analytic estimates predict an order-of-magnitude decrease in impurity sourcing and a 50% decrease in impurity radiation with SiC walls relative to graphite, assuming a very low C impurity fraction in the edge plasma can be achieved. This promising result motivates further investigations of silicon carbide as a low-Z, non-metallic PFM with favorable erosion properties. Long-range material migration of SiC needs still to be studied and understood to develop solutions to mitigate T retention by co-deposition, avoid dust formation/flaking from deposits, and prevent excessive dilution of the core plasma. The flexibility, diagnostic coverage, and rapid tile change-out capability of DIII-D makes it the ideal testbed to perform systematic evaluations of SiC material migration at a large scale. Sufficient coverage of the DIII-D first wall with SiC could have the added benefit of reducing the C impurity background in the plasma, allowing for more reactor-relevant studies of radiative optimization. High-fluence exposures of neutron-damaged SiC should also commence immediately on test stands and linear plamsa devices to evaluate the power handling and erosion properties of SiC in more detail; fortunately, initial effors are already underway [11][59].

With a coordinated research and development program it is estimated that the fission investment in silicon carbide could be leveraged to attain Technology Readiness Level (TRL) 4 in fusion applications within 10 years [60]. Rapid advances in industrial fabrication techniques, such as additive manufacturing and robotics, will be highly leveraging in this endeavor. Elevation of SiC PFC research to TRL4 or higher could be transformative to the worldwide PMI physics and fusion technology programs, placing SiC as a highly attractive design option for next-step devices.

**Acknowledgements**

This material is based upon work supported by the U.S. Department of Energy, Office of Science, Office of Fusion Energy Sciences, using the DIII-D National Fusion Facility, a DOE Office of Science user facility, under Award(s) DE-FC02-04ER54698. DIII-D data shown in this paper can be obtained in digital format by following the links at https://fusion.gat.com/global/D3D_DMP.

Disclaimer: This report was prepared as an account of work sponsored by an agency of the United States Government.

**References**

[1] Winter J 1996 *Plasma Phys. Control. Fusion* **38** 1503–1542
[2] Petty C C and The DIII-D Team 2019 *Nucl. Fusion* **59** 112002
[3] Doerner R P, Tynan G R and Schmid K 2019 *Nucl. Mater. Energy* **18** 56–61
[4] Stangeby P, Unterberg E A, Davis J W, Guo H Y, Rudakov D L, Leonard A W, Chrobak C P, Abrams T and Thomas D M "Developing refractory low-Z ceramic-clad plasma-facing components for robust burning plasma device operation," 2021 *Nucl. Fusion* in preparation
[5] Gomay Y, Konoshima S, Fujisawa N, Kasai S, Maeno M, Suzuki N, Hirayama T and Shimada M 1979 *Jpn. J. Appl. Phys.* 18 **7** 1318-1324
[6] Megusar J, Yang T F, Martin J R and Wan A S 1987 *Nucl. Instrum. Meth. B* **23** 476-481
[7] Li J, Wan B N, Luo J R, Kuang G L, Zhao Y P, Zhao J Y, Zhang X D, Liu X N, Fu P, Xie J K *et al* 2003 *Phys. Plasmas* **10** 1653
[8] Parker R, Janeschitz G, Pacher H D, Post D, Chiocchio S, Federici G, Ladd P, ITER Joint Central Team and Home Teams 1997 *J. Nucl. Mater.* **241-243** 1-26
[9] Khalifa H E, Koyanagi T, Jacobsen G M, Deck C P and Back C A 2017 *J. Nucl. Mater.* **487** 91-95
[10] Bringuier S, Abrams T, Guterl J, Vasudevamurthy G, Unterberg E, Rudakov D and Holland L 2019 *Nucl. Mater. Energy* **19** 1–6
[11] Koller M T, Davis J W, Goodland M E, Abrams T, Gonderman S, Herdrich G, Frieß M and Zuber C 2019 *Nucl. Mater. Energy* **20** 100704
[12] Rudakov D L, Wampler W R, Abrams T, Ding R, Boedo J A, Bringuier S, Bykov I, Chrobak C P, Elder J D and Guo H Y 2020 *Phys. Scripta* **T171** 014064
[13] Katoh K, Ozawa K, Shih C, Nozawa T, Shinavski R J, Hasegawa A and Snead L L 2014 *J. Nucl. Mater.* **448** 448-476
[14] Balden M, Picarle S and Roth J 2001 *J. Nucl. Mater.* **290-293** 47-51
[15] Westerhout J, Borodin D, Al R S, Brezinsek S,'t Hoen M H J, Kirschner A, Lisgo S, van der Meiden H J, Philipps V, van de Pol M J *et al* 2009 *Phys. Scr.* **T138** 014017
[16] Najmabadi F, Abdou A, Bromberg L, Brown T, Chan V C, Chu M C, Dahlgren F, El-Guebaly L, Heitzenroeder P, Henderson D *et al* 2006 *Fusion Eng. Des.* **80** 3–23
[17] Samm U, Bogen P, Esser G, Hey J D, Hintz E, Huber A, Könen L, Lie Y T, Mertens Ph, Philipps V *et al* 1995 *J. Nucl. Mater.* **220-222** 25-35
[18] Neu R, Kallenbach A, Krieger K, Rohde R and Roth J 2003 *Fusion Sci. Tech.* **44** 692-707
[19] Balden M and Roth J 2000 *J. Nucl. Mater.* **279** 351-355
[20] Sinclair G, Abrams T, Bringuier S, Doerner R P, Gonderman S, Holland L, Thomas D M and Yu J H "Quantifying erosion and retention of silicon carbide due to D plasma irradiation in a high-flux linear plasma device," 2021 *Nucl. Mater. Energy* submitted
[21] Mutzke A, Schneider R, Eckstein W and Dohmen R, "SDTrimSP: Version 5.00," 2011 *IPP, Report*
[22] Schmid K, Mayer M, Adelhelm C, Balden M, Lindig S and the ASDEX Upgrade team 2010 *Nucl. Fusion* **50** 105004
[23] Abe S, Skinner C H, Bykov I, Yeh Y W, Lasa A, Coburn J, Rudakov D L, Lasnier C J, Wang H Q, McLean A G *et al* "Measurement of Ion Impact Angle Distributions at Divertor Surfaces Using Micro-Engineered Targets on DiMES at DIII-D" 2021 *Nucl. Mater. Energy* submitted
[24] Abrams T, Ding R, Guo H Y, Thomas D M, Chrobak C P, Rudakov D L, McLean A G, Unterberg E A, Briesemeister A R, Stangeby P C *et al* 2017 *Nucl. Fusion* **57** 056034
[25] Ding R, Rudakov D L, Stangeby P C, Wampler W R, Abrams T, Brezinsek S, Briesemeister A, Bykov I, Chan V S, Chrobak C P *et al* 2017 *Nucl. Fusion* **57** 056016
[26] Tersoff J 1988 *Phys. Rev. B* **37** 6991
[27] Roth J 1999 *J. Nucl. Mater.* **266-269** 51-57
[28] Rudakov D L, Stangeby P C, Wong C P C, McLean A G, Wampler W R, Watkins J G, Boedo J A, Briesemeister A, Buchenauer D A, Chrobak C P *et al* 2015 *J. Nucl. Mater.* **463** 605–610
[29] Wong C P C, Junge R, Phelps R D, Politzer P, Puhn F, West W P, Bastasz R, Buchenauer D, Hsu W, Brooks J *et al* 1992 *J. Nucl. Mater.* **196-198** 871-875
[30] Snead L L, Nozawa T, Katoh Y, Byun T-S, Kondo S and D A Petti 2007 *J. Nucl. Mater.* **371** 329-377
[31] Deck C P, Khalifa H E, Sammula B, Hilsabeck T and Back C A 2012 *Prog. Nucl. Energ.* **57** 38-45
[32] Deck C P, Jacobsen G M, Sheeder J, Gutierrez O, Zhang J, Sonte J, Khalifa H E and Back C A 2015 *J. Nucl. Mater.* **466** 667-681
[33] Carlstrom T N, Hsieh C L and Stockdale R 1997 *Rev. Sci. Instrum.* **68** 1195
[34] Brooks N H, Howald A, Klepper K and West P 1992 *Rev. Sci. Instrum.* **63** 5167
[35] Bykov I, Chrobak C P, Abrams T, Rudakov D L, Unterberg E A, Wampler W R, Hollmann E M, Moyer R A, Boedo J A, Stahl B *et al* 2017 *Phys. Scr.* **T170** 014034
[36] Pospieszczyk A, Borodin D, Brezinsek S, Huber A, Kirschner A, Mertens Ph, Sergienko G, Schweer B, Beigman I L and Vainshtein L 2010 *J. Phys. B: At. Mol. Opt. Phys.* **43** 144017
[37] Bortolon A, Maingi R, Nagy A, Ren J, Duran J D, Maan A, Donovan D, Boedo J A, Rudakov D L, Hyatt A W *et al* 2020 *Nucl. Fusion* **60** 126010
[38] Summers H P, O'Mullane M G, Whiteford A D, Badnell N R and Loch S D 2007 *AIP Conf. Proc.* **901** 239-248

[39] Naujoks D, Asmussen K, Bessenrodt-Weberpals M, Deschka S, Dux R, Engelhardt W, Field A R, Fussmann F, Fuchs J C, C. Garcia-Rosales *et al* 1996 *Nucl. Fusion* **36** 671-687
[40] Abrams T, Thomas D M, Unterberg E A and Briesemeister A R *IEEE Trans. Plasma Sci*. **46** 1298-1305
[41] Brezinsek S, Pospieszczyk A, Borodin D, Stamp M F, Pugno R, McLean A G, Fantz U, Manhard A, Kallenbach A, Brooks N H *et al* 2007 *J. Nucl. Mater.* **363-365** 1119-1128
[42] Pugno R, Kallenbach A, Heger B, Carlson A and the ASDEX Upgrade Team 2003 *Proceedings of the 30th EPS Conference on Contr. Fusion and Plasma Phys.* **27A** P-1.153
[43] Nilson D G, Fenstermacher M E and Ellis R 1999 *Rev. Sci. Instrum.* **70** 738
[44] Stangeby P C 2000 The Plasma Boundary of Magnetic Fusion Devices (London: Taylor and Francis)
[45] Guillemaut C , Metzger C, Moulton D, Heinola K, O'Mullane M, Balboa I, Boom J, Matthews G F, S Silburn, Solano E R *et al* 2018 *Nucl. Fusion* **58** 066006
[46] Abrams T, Unterberg E A, Rudakov D L, Leonard A W, Schmitz O, Shiraki D, Baylor L R, Stangeby P C, Thomas D M and Wang H Q *Phys. Plasmas* 2019 **26** 062504
[47] Brezinsek S, Pugno R, Fantz U, Manhard A, Müller H W, Kallenbach A, Mertens Ph and the ASDEX Upgrade team 2007 *Phys. Scr.* **T128** 40-44
[48] Kawazome H, Ohya K, Inai K, Kawata J, Nishimura K and Tanabe T 2010 *Plasma Fusion Res.* **5** S2073
[49] McLean A G, Stangeby P C, Bray B D, Brezinsek S, Brooks N H, Davis J W, Isler R C, Kirschner A, Laengner R, Lasnier C J *et al* 2011 *J. Nucl. Mater.* **415** S141–S144
[50] Garçon M. Sauzéat L, Carlson R W, Shirey S B, Simon M, Balter V and Boyet M 2017 *Geostand. Geoanal. Res.* **41** 367-380
[51] Nieuwenhuizen P J 2001 *Appl. Catal. A-Gen.* **207** 55–68
[52] Jakubowski M W, Evans T E, Fenstermacher M E, Groth M, Lasnier C J, Leonard A W, Schmitz O, Watkins J G, Eich T, Fundamenski W *et al* 2009 *Nucl. Fusion* **49** 095013
[53] Eich T, Thomsen H, Fundamenski W, Arnoux G, Brezinsek S, Devaux S, Herrmann A, Jachmich A, Rapp J and JET-EFDA contributors 2011 *J. Nucl. Mater.* **415** S856-S859
[54] Pitts R A, Bonnin X, Escourbiac F, Frerichs H , Gunn J P, Hirai T, Kukushkin A S, Kaveeva E, Miller M A, Moulton D *et al* 2019 *Nucl. Mater. Energy* **20** 100696
[55] Behrisch R, Federici G, Kukushkin A and Reiter D 2003 *J. Nucl. Mater.* **313-316** 388-392
[56] Causey R A 2003 *J. Nucl. Mater.* **313-316** 450-454
[57] Roth J 1990 *J. Nucl. Mater.* **176-177** 132-141
[58] Plank H, Schwörer R and Roth J 1996 *Nucl. Instrum. Meth. B* **111** 63-69
[59] Beers C J, Lindquist E G, Biewer T M, Caneses J F, Caughman J B O, Goulding R H, Kafle N, Ray H, Showers M A, Zinkle S J *et al* 2019 *Fusion Eng. Des.* **138** 282-288
[60] Tillack M S and Holland L, "Accelerated Development of Silicon Carbide Composites for an Attractive Fusion Energy Source," 2017 *FESAC-TEC PMI Workshop*